**A primordial atmospheric origin of hydrospheric deuterium enrichment on Mars**

Kaveh Pahlevan[1,2]*, Laura Schaefer[3], Lindy Elkins-Tanton[1], Steven J. Desch[1], Peter R. Buseck[1]

1. School of Earth & Space Exploration, ASU, Tempe, AZ 85287, USA
2. Carl Sagan Center, SETI Institute, Mountain View, CA 94043, USA
3. Dept. of Geological Sciences, Stanford University, Stanford, CA 94305, USA

*To whom correspondence should be addressed: kaveh.pahlevan@asu.edu

**Abstract**

The deuterium-to-hydrogen (D/H or $^2$H/$^1$H) ratio of Martian atmospheric water (~6× standard mean ocean water, SMOW) is higher than that of known sources, requiring planetary enrichment. A recent measurement by NASA's Mars Science Laboratory rover *Curiosity* of Hesperian-era (>3 Ga) clays yields a D/H ratio ~3×SMOW, demonstrating that most of the enrichment occurs early in Mars's history, reinforcing the conclusions of Martian meteorite studies. As on Venus, Mars's D/H enrichment is widely thought to reflect preferential loss to space of $^1$H (protium) relative to $^2$H (deuterium), but both the cause and the global environmental context of large and early hydrogen losses remain to be determined. Here, we apply a recent model of primordial atmosphere evolution to Mars, link the magma ocean of the accretion epoch with a subsequent water-ocean epoch, and calculate the behavior of deuterium for comparison with the observed record. In contrast to earlier works that consider Martian D/H fractionation in atmospheres in which hydrogen reservoirs are present exclusively as $H_2O$ or $H_2$, here we consider 2-component ($H_2O$-$H_2$) outgassed atmospheres in which both condensed ($H_2O$) and escaping ($H_2$) components – and their interaction – are explicitly calculated. We find that a ≈2-3× hydrospheric deuterium-enrichment is produced rapidly if the Martian magma ocean is chemically reducing at last equilibration with the primordial atmosphere, making $H_2$ and CO the initially dominant species, with minor abundances of $H_2O$ and $CO_2$. Reducing gases – in particular $H_2$ – can cause substantial greenhouse warming and prevent a water ocean from freezing immediately after the magma ocean epoch. We find that greenhouse warming due to plausible $H_2$ inventories (p$H_2$=1-10$^2$ bars) yields surface temperatures high enough ($T_s$=290-560 K) to stabilize a water ocean and produce an early hydrological cycle through which surface water can be circulated. Moreover, the pressure-temperature conditions are high enough to produce ocean-atmosphere $H_2O$-$H_2$ isotopic equilibrium through gas-phase deuterium exchange such that surface $H_2O$ strongly concentrates deuterium relative to $H_2$, which preferentially takes up protium and escapes from the primordial atmosphere. The efficient physical separation of deuterium-rich ($H_2O$) and deuterium-poor ($H_2$) species via condensation permits equilibrium isotopic partitioning and early atmospheric escape to be recorded in modern crustal reservoirs. The proposed scenario of primordial $H_2$-CO-rich outgassing and escape suggests significant durations (>Myr) of chemical conditions on the Martian surface conducive to prebiotic chemistry immediately following magma ocean crystallization.

# 1. Introduction

The primordial Martian atmosphere is of major interest. Despite calculations indicating that most hydrogen and carbon outgas from the Martian magma ocean (Elkins-Tanton, 2008), the oxygen fugacity ($fO_2$) characterizing outgassing – and therefore the chemical composition of the resulting primordial atmosphere – remains unconstrained (Hirschmann, 2012). In particular, both oxidizing ($H_2O$-$CO_2$-rich) and reducing ($H_2$-CO-rich) primordial atmospheres have been advocated recently for early Mars (Cannon et al., 2017; Saito and Kuramoto, 2018). These end-member atmospheres have contrasting consequences for the primordial Martian climate. Whereas an $H_2O$-$CO_2$-rich atmosphere would condense into ice layers, producing an early icehouse (Kasting, 1991; Wordsworth et al., 2013), an $H_2$-CO-rich atmosphere is not condensable at Martian temperatures and would generate greenhouse warming and stabilize an ocean against freezing. An $H_2$-CO-rich Martian atmosphere would produce habitable surface conditions before $H_2$ exhaustion via escape, a possibility also proposed for young exoplanets (Pierrehumbert and Gaidos, 2011; Wordsworth, 2012). Despite decades of debate about these dramatically distinct early atmospheres and the associated environments, it remains to be clearly established whether Mars ever had an $H_2$-rich atmosphere (Dreibus and Wanke, 1985).

An approximately six-fold deuterium-to-hydrogen (D/H or $^2H/^1H$) enrichment in Martian water vapor relative to standard mean ocean water (SMOW, D/H=$1.56 \times 10^{-4}$) was first detected spectroscopically (Owen et al., 1988) and has since been confirmed by in-situ measurements of Mars's atmosphere (Webster et al., 2013). A key observation constraining Martian hydrogen history is that much – and perhaps most – of the hydrogen loss occurs in the first 500 Myr of Mars's history (Kurokawa et al., 2014). Because Mars does not experience plate tectonics and associated crustal recycling and volatile subduction, mantle water sampled via partial melts is thought to preserve the initial D/H of water accreted to Mars (Hallis et al., 2012). Recent inferences of the Martian mantle composition have largely converged to D/H values similar to carbonaceous chondrites or terrestrial ocean water (Greenwood et al., 2018; Hallis et al., 2012; Peslier et al., 2019; Usui et al., 2012). In the context of the Martian magma ocean, this is the inferred initial D/H of hydrogen (as $H_2$, $CH_4$, and $H_2O$) outgassed into the primordial atmosphere at 4.5 Ga (Elkins-Tanton, 2008). By contrast, a recent *Curiosity* rover measurement of Hesperian-era clays (~3.6 Ga) yields a hydrospheric D/H value ~3×SMOW (Mahaffy et al., 2015), revealing that most Martian deuterium enrichment occurs early in planetary history, reinforcing the conclusion of studies of aqueous alteration products in ~4 Ga ALH84001 (Boctor et al., 2003; Greenwood et al., 2008). Despite an increasingly broad sampling of the Martian hydrogen isotopic record (Usui, 2019), and recognition of the importance of $H_2$ to climate history (Ramirez et al., 2014; Wordsworth et al., 2017), the ≈2-3× deuterium-enrichment of the Martian hydrosphere in the first ~500 Myr of planetary history has yet to be explained in any model. Thus, the sequence of events that enrich early Martian waters in deuterium (Fig. 1) remains to be understood.

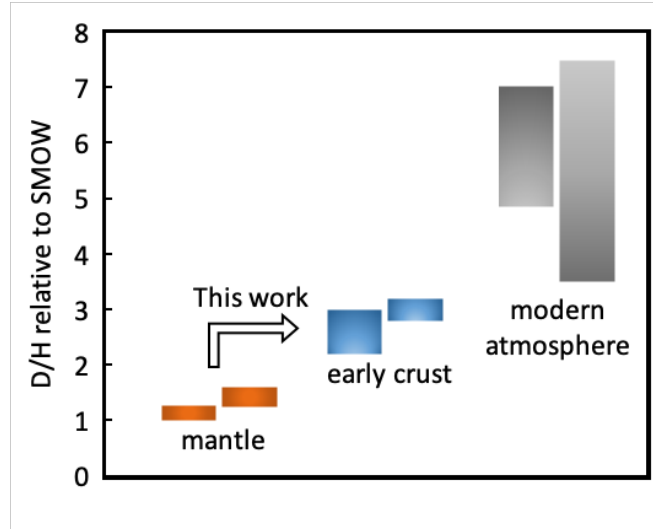

**Fig. 1. The Martian hydrogen isotopic record.** The depleted mantle has the lowest inferred D/H among the Martian reservoirs (Peslier et al., 2019; Usui et al., 2012), whose composition (≈SMOW) is consistent with chondritic inheritance without contamination by crustal deuterium-enriched sources. Intermediate values (≈2-3 × SMOW) are inferred for ~4 Ga carbonates in crustal sample ALH84001 (Boctor et al., 2003; Greenwood et al., 2008) and a ~3.6 Ga surface clay (Mahaffy et al., 2015). Such an intermediate reservoir is inferred to be global (Usui et al., 2015). Modern atmospheric water displays the highest D/H among all Martian reservoirs (Webster et al., 2013), presumably reflecting enrichment via preferential protium escape. Here, we constrain earliest Martian evolution using the D/H offset between mantle and crust/hydrosphere as a constraint.

Although the preferential loss to space of $^1$H (protium) relative to $^2$H (deuterium) is widely accepted as the origin of the Martian hydrospheric deuterium enrichment, the environmental and evolutionary context of a massive early loss of hydrogen remains to be determined. In particular, modern hydrogen loss rates due to photodissociation of water (Yung et al., 1988) are several orders of magnitude too low to isotopically enrich a massive early hydrosphere within the available time (Scheller et al., 2021). Hydrodynamic escape from an early $H_2$-rich atmosphere can produce larger escape rates such that $H_2$/HD mass fractionation could potentially enrich the *atmospheric* $H_2$ inventory in deuterium (Zahnle et al., 1990). As an explanation for the observed deuterium enrichment, however, this process faces problems. First, a sufficiently vigorous $H_2$ wind induces negligible mass fractionation as deuterium is swept along with the escaping hydrogen via molecular collisions (Genda and Ikoma, 2008). Second, signatures of Martian deuterium enrichment are expressed in surface *aqueous* alteration products (Boctor et al., 2003; Greenwood et al., 2008; Mahaffy et al., 2015), and it is not clear how the deuterium enrichment of atmospheric $H_2$ would be transmitted to hydrospheric $H_2O$ and thereby to hydrous minerals, where it is observed. Any viable scenario of Martian D/H enrichment must produce atmospheric hydrogen loss that is sufficiently vigorous *and* selective isotopically, with the deuterium enrichment signature transmitted to the surface hydrosphere where it is documented in the mineral record. Existing models of Martian deuterium enrichment either produce insufficient hydrogen loss rates (Yung et al., 1988) or neglect the transmission of the deuterium enrichment signature from atmospheric $H_2$ to hydrospheric $H_2O$ (Zahnle et al., 1990). Interpretation of the observed hydrogen isotope record of Mars thus requires a consistent scenario of hydrospheric and atmospheric evolution.

A useful point of comparison is the history of the terrestrial hydrosphere, which has experienced minimal D/H enrichment (<a few percent) over the past 3.8 Ga (Pope et al., 2012). The observed constancy of terrestrial D/H over time can be understood by examining the modern Earth: water vapor is depleted down to the ppm level in the stratosphere due to condensation in the upper troposphere (~190-210 K, the "cold-trap"), limiting the hydrogen flux to the upper atmosphere from where escape can occur (Hunten, 1993). At the modern rate of escape, the amount of hydrogen lost from Earth over geologic time comprises a ~2 m global equivalent layer (GEL) of water. Given the mass of Earth's hydrosphere (>3 km GEL), such escape rates cause only negligible D/H enrichment, even if deuterium is entirely retained during the escape process. It is recognized that non-condensable gases (e.g., $CH_4$) traverse the cold-trap unimpeded and enhance hydrogen loss rates (Catling et al., 2001; Zahnle et al., 2019), but the relative constancy of terrestrial D/H places limits on the extent of such reduced gas loss during the Archean (Kurokawa et al., 2018) and Hadean (Pahlevan et al., 2019) eons. In the Martian case, hydrospheric deuterium enrichment implying massive hydrogen loss is documented in the volatile record but remains poorly understood. Some physical process apparently produces a recognizable deuterium enrichment in the hydrosphere of Mars but not Earth.

Here, we propose the hypothesis that Mars forms with a chemically reducing magma ocean that coexists with a primordial reducing ($H_2$-CO-rich) atmosphere upon crystallization and that this atmosphere determines the earliest Martian climate, escape rates, and isotopic fractionation. Such a reducing atmosphere is in contrast to the oxidizing ($H_2O$-$CO_2$-rich) atmosphere considered likely for magma ocean outgassing on Earth (Armstrong et al., 2019; Hirschmann, 2012)[1].

In contrast to earlier works that consider Martian D/H fractionation in atmospheres where hydrogen reservoirs are present exclusively as $H_2O$ (Yung et al., 1988) or $H_2$ (Zahnle et al., 1990), here we consider two-component ($H_2O$-$H_2$) outgassed atmospheres in which both condensed ($H_2O$) and escaping ($H_2$) components – and their interaction – are explicitly calculated. Despite uncertainties about initial gas inventories, accreted volatile elements are apparently subject to high-temperature processing with a Martian magma ocean (Elkins-Tanton, 2008), a process that likely resets the volatile speciation. The initial conditions for Martian atmospheric and hydrospheric evolution are therefore likely established by high-temperature thermodynamics in a steam atmosphere, which can be calculated. We propose that equilibrium isotopic partitioning between $H_2O$ and $H_2$ is the dominant process that determines the early deuterium enrichment of the Martian hydrosphere, an enrichment that depends on the outgassed $H_2/H_2O$ ratio and thus the oxygen fugacity at the magma ocean surface. We thereby propose that the Martian D/H record yields an oxybarometer for the primordial atmosphere, and that the observed ≈2-3× early enrichment reflects the reducing conditions characterizing last equilibration with a Martian magma ocean. We describe an atmospheric model linked to – and constrained by – the hydrogen isotopic record to characterize an early $H_2$-rich epoch in Martian history immediately following the magma ocean produced during accretion.

---

[1] Oxidizing in this context refers to the silicate mantle $Fe^{3+}/Fe^{2+}$ ratio, which determines magmatic $fO_2$ and the gas phase speciation, in contrast to discussions of building blocks for the bulk planet in which an oxidizing component is associated with higher mantle FeO, which leads to the statement that Mars is oxidized (Dreibus and Wanke, 1985). In the context of silicate $Fe^{3+}/Fe^{2+}$, the Martian mantle is more reduced than Earth's.

The outline of the paper is: §2 – summary of the model used to describe primordial atmosphere evolution during the magma ocean and early water ocean epochs and the associated D/H signatures, §3 – results for earliest Martian climate and the character of escape, as well as hydrogen isotopic oxybarometry of the primordial atmosphere, §4 – discussion of the removal of the primordial atmosphere, the later D/H evolution, and the connection with the redox state of silicate Mars, and §5 – summary and conclusions.

## 2. Atmosphere evolution model

Mars and its primordial atmosphere are thought to be coeval, forming via the same process: impacts of solid bodies with proto-Mars during accretion (Saito and Kuramoto, 2018). Experiments on silicate materials demonstrate that once proto-Mars reaches ~0.1 Mars masses ($M_M$), shock-heating of accreting materials generates post-impact temperatures high enough to destabilize hydrous silicates, leading to degassing of impactors and deposition of volatiles into the primordial atmosphere (Tyburczy et al., 1986). In this section, we describe the model used to characterize the primordial atmosphere during (§2.1) and after (§2.2) Martian accretion. We then describe deuterium as a tracer of the primordial atmosphere using equilibrium isotopic partitioning (§2.3). By coupling the origin and early evolution of Mars with that of its atmosphere and hydrosphere, we describe the first model that can explain – and be constrained by – the early D/H enrichment observed in the Martian volatile record.

### 2.1. Magma ocean outgassing

The chemical composition of the primordial Martian atmosphere – especially the oxygen fugacity that determines its $H_2/H_2O$ ratio – is thought to be determined via interaction with an underlying magma ocean. Several lines of evidence suggest that a magma ocean forms during Martian accretion, including trace element signatures of silicate differentiation in the SNC meteorites (Elkins-Tanton et al., 2005), short-lived radionuclide evidence of volatile/refractory element separation in the first 30 Myr of Mars history (Marty and Marti, 2002), and core-mantle differentiation in the first 2-4 Myr of the Solar System (Dauphas and Pourmand, 2011), a process that requires large-scale melting. Indeed, the thermal-blanketing effect of an impact-degassed atmosphere plays a critical role in retaining the heat of accretion and facilitating the formation of a magma ocean (Saito and Kuramoto, 2018). Thus, the primordial Martian atmosphere likely coexists with a magma ocean (Elkins-Tanton, 2008).

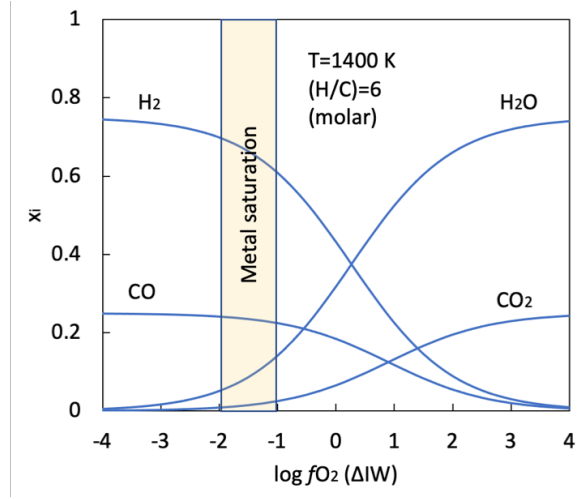

**Fig. 2. Primordial atmosphere composition produced via equilibrium with a magma ocean.** The mole fraction of atmospheric vapor species is calculated as a function of the oxygen fugacity ($fO_2$) at an equilibrium temperature of 1,400 K. Parameters for the IW buffer are given in (Frost, 1991) and thermodynamic data for gaseous species ($H_2O$-$H_2$-$CO$-$CO_2$) are adopted from (Chase et al., 1985). Magma oceans with suspended metallic droplets have $fO_2$ buffered to a particular value, which, for the Martian mantle FeO content, is equal to ΔIW=-1 to -2 (Brennan et al., 2022) ("Metal saturation"). Figure adapted from (Pahlevan et al., 2019).

A remarkable feature of high-temperature thermodynamics of magmatic gas mixtures is that they can be dominated by $H_2$-CO or by $H_2O$-$CO_2$, depending on the oxygen fugacity of the magma ocean with which they equilibrate (Fig. 2). Recent estimates of the timescales for magma ocean-atmosphere equilibration ($\approx 10^3$-$10^4$ years) suggest that this process is rapid relative to evolutionary timescales ($\approx 10^6$-$10^7$ years) (Hamano et al., 2013; Pahlevan et al., 2019). In principle, one could calculate the $H_2/H_2O$ ratio of the primordial atmosphere from the $fO_2$ recorded in magma ocean crystallization products sampled via melts from the Martian mantle; however, the inferred $fO_2$ of source regions of the Martian mantle range from chemically reducing (ΔIW≈0) to oxidizing (ΔIW≈+4) conditions (Castle and Herd, 2017; Nakada et al., 2020; Wadhwa, 2008) (ΔIW denotes the $\log_{10}$ deviation of $fO_2$ from the iron-wüstite buffer). We therefore adopt a two-component ($H_2O$-$H_2$) model to describe the primordial atmosphere, with oxygen fugacity (or equivalently $H_2/H_2O$ ratio) of the outgassed mixture a free parameter whose value must be constrained by hydrogen isotopic oxybarometry (§2.3). The connection with the redox state of the Martian mantle is discussed later (§4.3).

*2.2. Primordial climate*
The primary control on early climates is the presence of an ocean and the atmospheric $H_2$ inventory. When the heat input from accretionary impacts falls below a certain threshold, the Martian magma ocean cools and crystallizes, expelling most of its dissolved water into the atmosphere. This expulsion leads to rapid ocean condensation on ~$10^3$ year timescales (Abe, 1993). Following ocean formation, which removes most outgassed water vapor, a dense atmosphere composed of gases with low aqueous solubility remains (see supplementary §A1 for a discussion of atmospheric stability against rapid blow-off). We expect $H_2$ to play a dominant role in determining the equilibrium climate in such an atmosphere for several reasons. First, even in moderately reducing outgassed atmospheres (e.g., ΔIW≈0), $H_2$ molecules are abundant by number (Fig. 2). Second, whereas a pure $CO_2$-based greenhouse cannot stabilize an ocean on Mars due to the limiting effects of condensation (Kasting, 1991; Wordsworth et al., 2013), an $H_2$-based greenhouse does not condense at planetary temperatures and can stabilize a water ocean against freezing until it is stripped away by radiation or impacts (see §4.1 for a discussion of atmospheric removal). Collision-induced infrared opacity makes $H_2$ a potent greenhouse gas at pressures ≳1 bar (Pierrehumbert and Gaidos, 2011; Sagan, 1977; Sagan and Mullen, 1972; Wordsworth, 2012). The existence of such opacity has led to proposals of habitable conditions on young exoplanets beyond the traditional habitable zone (Pierrehumbert and Gaidos, 2011; Wordsworth, 2012). The aqueous solubility of $H_2$ is low (Gordon et al., 1977), facilitating its availability as an atmospheric greenhouse gas. Finally, because methane is unstable with respect to photodissociation via solar Lyman α photons (Kasting, 2014), the dominant reducing hydrogen-bearing gas in primordial atmospheres – and the carrier molecule for escaping hydrogen – is expected to be $H_2$ (Zahnle et al., 2020).

We calculate the climate consequences of a primordial Martian atmosphere, adopting the two-component ($H_2O$-$H_2$) chemical model from §2.1, which we use to characterize greenhouse warming via $H_2$ opacity and the moderation of surface temperatures via tropospheric condensation. We consider one-dimensional thermal and chemical vertical structure by calculating moist adiabats in the all-troposphere approximation, iterating to find solutions for surface temperature (Pierrehumbert, 2010). Due to the condensation of water in the lower atmosphere and collision-induced infrared opacity of $H_2$ at moderate (~0.1 bar) pressures, we expect the infrared opacity of $H_2$ to determine the thermal emission level. The effective temperature ($T_E$) is given by top-of-the-atmosphere radiative balance with the young Sun:

$$\frac{L}{4}(1-A) = \sigma_{SB} T_E^4 \quad (1)$$

with L the solar constant for early Mars, A the bond albedo, and $\sigma_{SB}$ the Stefan-Boltzmann constant. For L=440 W/m² appropriate for the early Sun (with 0.75 × the present-day luminosity) at 1.52 AU and a bond albedo A=0.25 similar to that of Uranus or Neptune (Lodders and Fegley, 1998), the effective temperature for early Mars is ~195 K. Calculated results are qualitatively similar for somewhat different effective temperatures (175-210 K), as expected from cloud feedback or adjustments to planetary albedo that account for Rayleigh scattering in thicker atmospheres, effects known to alter the radiation budget by tens of percent (Wordsworth, 2012). At infrared wavelengths, an optical depth unity surface of a pure $H_2$ atmosphere of a cool ($T_E$=100 K) super-Earth (g=20 m/s²) is ~0.2 bars (Wordsworth, 2012). Scaling the photospheric pressure to the effective temperature ($T_E$=195 K) and gravity (g=3.75 m/s²) relevant to early Mars yields an emission level pressure of ~0.12 bars. Because the atmosphere at the emission level is cold and dry [$p_{H_2O}$- $_{sat}$(195K)≈$10^{-6}$ bars], we take this pressure – appropriate for a pure $H_2$ atmosphere – as the emission-level pressure to which a moist $H_2O$-$H_2$ adiabatic structure must be stitched.

At the bottom boundary of the atmosphere, two-phase equilibrium with the ocean dictates water vapor abundance. We assume a troposphere saturated in water vapor throughout. The evolution of primordial climate in this model is thus characterized by one free parameter: the surface pressure of an equivalent pure $H_2$ inventory (p$H_2$), which is initially endowed and gradually depletes via escape. The procedure used to calculate surface temperatures is described in detail elsewhere (Pahlevan et al., 2019). The results of the climate calculations are then used to describe the isotopic evolution of the Martian hydrosphere with which the atmosphere is in contact (§2.3).

*2.3. Hydrogen isotopic evolution of the Martian hydrosphere*
In our model, the deuterium content of the earliest Martian hydrosphere is established by equilibrium partitioning. After magma ocean solidification, geothermal heat flow becomes climatologically insignificant and surface temperatures drop precipitously, from >1,500 K (Saito and Kuramoto, 2018) to <600 K (Abe, 1993), being determined by solar radiative balance alone. Such a massive temperature drop leads to dramatically lower rates of isotopic exchange among gaseous species such that all regions of the atmosphere become quenched with respect to $H_2O$-$H_2$ deuterium exchange on the timescales characterizing atmospheric mixing. In this quenched isotopic exchange regime, the atmosphere and hydrosphere are both internally well-mixed at any given epoch, and each reservoir is nominally characterized by a single value of D/H (Genda and Ikoma, 2008; Pahlevan et al., 2019).

In contrast to the behavior of deuterium on giant planets, no rapid high-temperature exchange occurs at atmospheric depths in this exchange regime, and another timescale – the residence time of $H_2$ with respect to atmospheric escape – determines whether ocean-atmosphere isotopic equilibration is achieved. In a multi-bar $H_2$-rich atmosphere, the timescale for ocean-atmosphere hydrogen isotopic equilibration (T$_E$) through gas-phase exchange is rapid (≈$10^3$-$10^6$ years) relative to the timescales for atmospheric $H_2$ escape (e.g., ≈$10^7$-$10^8$ years, Fig. A1) ensuring continuous equilibration during the loss process (Genda and Ikoma, 2008; Pahlevan et al., 2019).

In this quasi-equilibrium state at each epoch, the magnitude of hydrospheric deuterium enrichment depends on climate. As the ocean and atmosphere settle into a solar-powered climate with a hydrological cycle, evaporation of the ocean water and the general circulation of the atmosphere provide an opportunity for $H_2O$-$H_2$ isotopic equilibration. Equilibrium between the ocean and the $H_2$-rich atmosphere initially concentrates deuterium in the ocean, with the magnitude of the enrichment dependent not only on the relative size of the atmospheric $H_2$ and hydrospheric $H_2O$ reservoirs, but also the temperature of equilibration. This equilibrium can be described with an exchange reaction and an associated temperature-dependent equilibrium constant:

$$H_2O + HD \Leftrightarrow HDO + H_2 \qquad (2)$$
$$K_{EQ} = 1 + 0.22 \times (10^3/T)^2 \qquad (3)$$

with the latter expression accurate to good (~10%) approximation (Richet et al., 1977). Although this equilibrium formally applies to gaseous molecules, it can be used to describe $H_2O$-$H_2$ ocean-atmosphere equilibrium because the vapor pressure isotope effect for the phase equilibrium reaction [$H_2O$(liquid)⇔$H_2O$(vapor)] is at least an order-of-magnitude smaller and can be neglected to adequate approximation, such that water vapor reflects the isotopic composition of the ocean and the rest of the hydrosphere generally. Although in principle hydrogen isotopic exchange occurs throughout the atmospheric column, in practice the strong temperature-dependence of the reaction kinetics (Lécluse and Robert, 1994) makes the basal atmosphere the relevant environment for ocean-atmosphere equilibration. Calculation of the hydrogen isotopic evolution of the ocean requires knowing the temperature at the interface with the $H_2$-rich atmosphere, for which we adopt the surface temperatures from the climate model (§2.2). For the range of surface temperatures of interest (≈300-600 K, see §3.1), the variation in the equilibrium constant is substantial ($K_{EQ}$=1.6-3.4). An accurate climate model is therefore required for an accurate description of the hydrosphere-atmosphere isotopic fractionation.

Isotopic evolution of the Martian hydrosphere is calculated as a series of equilibrium steps. Because deuterium is a trace constituent in all such scenarios ([D/H]~$10^{-4}$), the equilibrium constant can be rewritten (Genda and Ikoma, 2008) and supplemented with an isotopic mass-balance equation:

$$K_{EQ} \approx D_{H_2O}/D_{H_2} = 1 + 0.22(10^3/T)^2 \qquad (4)$$
$$D_T = F_{H_2O}D_{H_2O} + F_{H_2}D_{H_2} \qquad (5)$$

where $D_{H_2O}$, $D_{H_2}$, and $D_T$ are the D/H ratios in the oceans, the atmosphere, and fluid envelope as a whole (ocean plus atmosphere), $F_{H_2O}[=N_{H_2O}/(N_{H_2O}+N_{H_2})]$ and $F_{H_2}[=N_{H_2}/(N_{H_2O}+N_{H_2})]$ represent the molar fraction of total fluid envelope hydrogen in the hydrosphere and atmosphere, respectively, and $N_{H_2O}$ and $N_{H_2}$ are the total number of water and hydrogen molecules in the fluid envelope, respectively. We neglect the possible role of methane because – if present – it rapidly photodissociates and converts to $H_2$ before escape, as also occurs on the early Earth (Kasting, 2014; Zahnle et al., 2020).

At each step of the calculation, the procedure for determining hydrospheric isotopic evolution involves: (1) calculation of surface temperature for a given greenhouse inventory [$T(pH_2)$], which, along with the relative mass of the ocean and atmosphere, determines the isotopic partitioning upon equilibration; (2) non-fractionating removal (Yoshida and Kuramoto, 2020)[2] of some atmospheric $H_2$ ($N_{H_2}$), diminishing greenhouse warming and lowering surface temperatures (Fig. 3); and (3) ocean-atmosphere re-equilibration at a new surface temperature, accentuating the H isotopic contrast between the reservoirs by elevating the D/H of the hydrosphere and lowering that of atmospheric $H_2$ (see Eqn. 3). We use the approximation that $H_2O$ condenses in the troposphere and is retained whereas $H_2$ is transported to the upper atmosphere and lost (see §3.2 and Fig. 4 for justification). In this way, even though primordial Martian water condenses into an ocean and clouds and is entirely retained in the model (similar to the terrestrial hydrosphere), its deuterium content can evolve via equilibrium exchange with an unbound atmospheric $H_2$ reservoir. We calculate this compositional evolution until the exhaustion of the primordial $H_2$ inventory, at which point we report hydrospheric D/H compositions (Fig. 5).

---

[2] Non-fractionating removal of an $H_2$-rich atmosphere potentially underestimates the D/H enrichment, because we ignore kinetic HD/$H_2$ mass-fractionation that accompanies the hydrodynamic escape process itself. Recent calculations, however, suggest that the kinetic mass-fractionation effect for hydrogen is minor and subdominant to the equilibrium isotope effect we consider.

## 3. Results
### 3.1. Primordial climate
We find the greenhouse effect accompanying a multi-bar $H_2$ atmosphere is sufficient to stabilize a surface ocean climate on primordial Mars. Calculated surface temperatures for an $H_2$ Martian greenhouse co-existing with a water ocean depend primarily on the molecular hydrogen inventory ($\propto pH_2$), whereas the mass of atmospheric water vapor is determined – as on the Earth – via vapor pressure equilibrium with the underlying ocean. Primordial inventories equivalent to $1$-$10^2$ bars of pure $H_2$ stabilize a water ocean climate at surface temperatures 290-560 K (Fig 3). The existence of such greenhouse solutions makes possible a water ocean epoch on earliest Mars whose duration depends on the thickness of the $H_2$-rich atmosphere, which we later constrain using deuterium (§3.3).

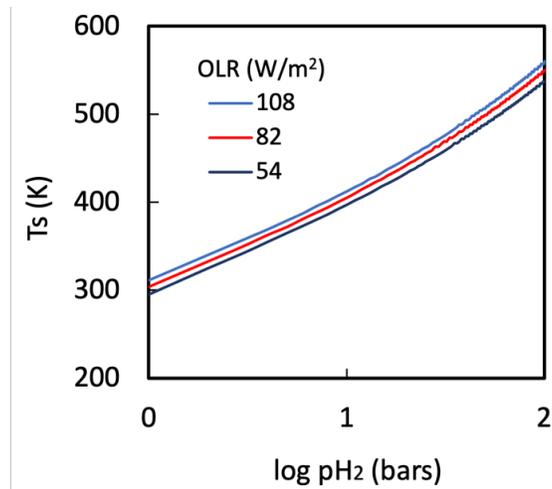

**Fig. 3. Surface temperature as a function of primordial Martian $H_2$ inventory.** Partial pressure is used as a proxy for the size of the gaseous inventory, and is defined as equivalent pressure of a pure $H_2$ atmosphere ($pH_2=\sigma H_2 g$), which determines the magnitude of the greenhouse effect. Outgoing longwave radiation (OLR) corresponds to effective temperatures (176, 195, 209 K) that encompass a plausible range of planetary bond albedos (0-0.5). The main effect of tropospheric water vapor is to stabilize water oceans by lowering surface temperatures by reducing the adiabatic lapse rate from the photosphere to the surface from dry to moist. Earliest Martian climate in these scenarios depends mainly on the $H_2$ inventory.

*3.2. Divergent fates for $H_2O$ and $H_2$*

In the isotopic evolution model (§2.3), we adopt the approximation that outgassed $H_2O$ is retained via condensation into oceans, lakes, and clouds on or near the planetary surface, whereas outgassed $H_2$ is lost via atmospheric escape. The closed system behavior of $H_2O$ can be evaluated for two sequential epochs, that of: (i) the magma ocean and (ii) the solar-powered water ocean.

During the magma ocean epoch, there is no planetary hydrosphere (except for perhaps condensation clouds) and all gases in the atmosphere are potentially subject to loss. However, the atmosphere in this epoch is expected to be fully convective (Salvador et al., 2017), and any chemical evolution due to gas loss is quickly communicated down to the magma ocean-atmosphere interface, influencing the $fO_2$ of magma-vapor equilibrium. For this reason, characterizing the chemical composition of the atmosphere via the oxygen fugacity of last equilibration (e.g., Fig. 2) efficiently captures the early evolution of the magma ocean-atmosphere system, including open system losses. Once the magma ocean crystallizes at the surface, the steam atmosphere condenses into a surface hydrosphere on a rapid timescale (~$10^3$ years) relative to escape timescales (Abe, 1993), such that water loss during the transition to a solar-powered climate is also expected to be negligible.

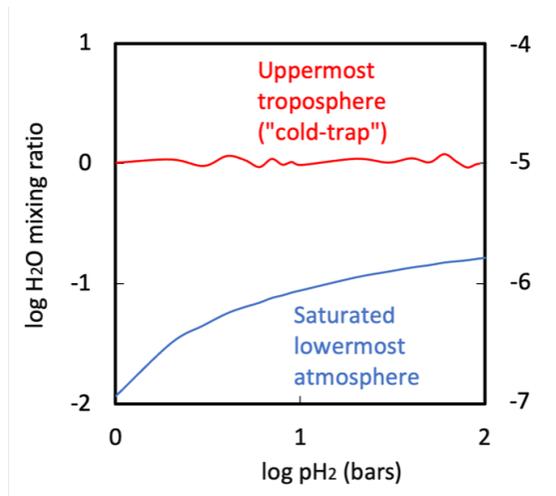

**Fig. 4. Mixing ratio of water vapor in the atmosphere at ocean surface (blue, left axis) and emission level (red, right axis).** $H_2O$-$H_2$ atmospheres feature a troposphere that is moist but a stratosphere that is dry due to a condensation cold-trap as on modern Earth. A constant [$H_2O$] in the upper troposphere results in this regime because the opacity at the emission level (p≈0.12 bars) is dominated by molecular hydrogen and the temperature at the emission level is given by top-of-the-atmosphere radiative balance, dictating low $H_2O$ vapor pressure via 2-phase equilibrium [$p_{H_2O\text{-sat}}$(195K)≈$10^{-6}$ bars]. The troposphere is assumed saturated in water vapor, yielding an upper limit to stratospheric $H_2O$ mixing ratios which are nevertheless quite small (<10 ppmv).

During the ensuing solar-powered water ocean epoch, water vapor is present in relatively moderate abundances (≈1-10% by number) in the lower atmosphere due to vapor pressure equilibrium with surface oceans (blue curve, Fig 4). These abundances are low enough (Kasting, 1988) such that the bulk of the vapor is removed via condensation below the tropopause, producing a very dry stratosphere (<10ppmv $H_2O$, red curve, Fig 4). The existence of these efficient cold-trap structure solutions supports the assertion that $H_2O$ is retained in the troposphere via condensation, whereas $H_2$ is transported to the upper atmosphere and lost via exposure to high-energy solar radiation.

### 3.3. D/H oxybarometry

Using the results of the climate model (§3.1) and the divergent fate approximation (§3.2), we calculate the isotopic evolution of the hydrosphere for various conditions of primordial outgassing. Each value of magma ocean oxygen fugacity yields a unique value for the outgassed $H_2/H_2O$ ratio. In order to convert this ratio into an initial atmospheric inventory ($pH_2$), we scale by the reservoir size of the early Martian hydrosphere, which has been estimated as ≈500 m GEL (Di Achille and Hynek, 2010) and is constant in our model. This hydrospheric reservoir is equivent to ≈18 bars of $H_2O$ pressure on the Martian surface and contains the equivalent of ≈2 bars of $H_2$, which serves as a reference scale for isotopic mass balance involving the atmospheric $H_2$ inventory (see §2.3 for details).

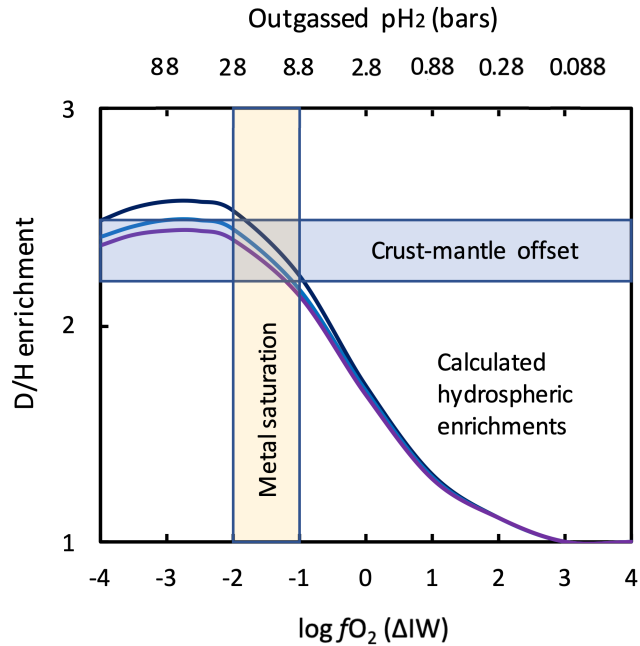

**Fig. 5. Hydrospheric D/H enrichment versus oxygen fugacity of primordial outgassing.** Oxygen fugacity determines $H_2/H_2O$ of primordial outgassing, which can be expressed as $pH_2$ by scaling by the size of the early hydrosphere. High $fO_2$ (oxidizing) outgassing leads to nearly "pure" steam atmospheres with minimal $H_2$ escape and deuterium-enrichment of the hydrosphere, whereas more reducing conditions (ΔIW<0) generate

higher $H_2$ abundances in the outgassed envelope ($H_2/H_2O>1$) and stronger deuterium-enrichments ($f^{D/H}>2$) due to isotopic equilibration and removal of isotopically light $H_2$. The three enrichment curves correspond to different emission temperatures ($T_E$=176, 195, 209 K) corresponding to different planetary albedos, demonstrating the robustness of the result to plausible variations in early climate. The inferred $fO_2$ of the Martian magma ocean buffered by the presence of metallic iron during core formation yields an outgassed composition and hydrospheric deuterium-enrichment ($\approx$2-3×) that overlaps with the crust-mantle D/H offset observed in the volatile record (see Fig. 1).

We find that early hydrospheric D/H enrichment ($f^{D/H}\equiv(D/H)_{H_2O}/(D/H)_{initial}$) is primarily a function of the oxygen fugacity of primordial outgassing (Fig. 5). Oxidizing conditions for outgassing ($\Delta IW>+1$) produce nearly pure steam atmospheres ($H_2/H_2O<1/3$, $pH_2<0.7$ bars) and thus minimal hydrospheric D/H enrichment ($f^{D/H}\approx1$) because only a small fraction of the total outgassed hydrogen appears as $H_2$ and escapes. Mildly reducing conditions for outgassing ($\Delta IW\approx0$) result in higher hydrospheric D/H enrichments due to a higher abundance of primordial $H_2$ molecules, which become deuterium-depleted and lost. For even more reducing compositions ($\Delta IW<-1$, $pH_2>10$ bars), the hydrospheric D/H enrichment reaches a near-constant plateau ($f^{D/H}\approx2.5$) as the greater leverage afforded by the larger $H_2$ inventory is counteracted by the higher surface temperatures arising from more greenhouse warming, the latter effect diminishing equilibrium $H_2O$-$H_2$ fractionation. Because final hydrospheric D/H is a strong function of the $fO_2$ of outgassing, we conclude that the D/H recorded by crustal aqueous alteration products in contact with Mars's early hydrosphere is a powerful oxybarometer for earliest conditions on the Martian surface, and a barometer for the $H_2$ abundance within the context of an $H_2$-$H_2O$ model atmosphere.

The past presence of an $H_2$-rich Martian atmosphere is supported by the finding that primordial outgassing at magma ocean redox conditions expected during Martian core formation (i.e., $\Delta IW<-1$) (Brennan et al., 2022) yields an outgassed molecular composition ($H_2/H_2O>4$, $pH_2>10$ bars) that robustly predicts a $\approx$2-3× D/H enrichment for the Martian hydrosphere relative to the Martian mantle (Fig. 5), an enrichment that reproduces the crust-mantle D/H offset observed in the Martian volatile record ($f^{D/H}\approx2.2$-2.5, Fig. 1) (Kurokawa et al., 2014; Usui et al., 2015). We conclude that primordial outgassing on Mars occurs at reducing conditions with abundant primordial $H_2$ ($pH_2>10$ bars) and discuss the connection with the redox state of the Martian mantle later (§4.3). A testable prediction of this scenario is that the Martian hydrosphere is deuterium-enriched from the epoch immediately after the magma ocean and the surface hydrosphere never displays the chondritic isotopic composition observed in the Martian mantle or the terrestrial oceans (Fig. 1). Better characterization of the D/H composition of the Noachian hydrosphere – recognized as a high-priority science goal (Usui, 2019) – subjects this prediction to an observational test.

## 4. Discussion

The chemical composition of the primordial Martian atmosphere has been underconstrained. In particular, both oxidizing ($H_2O$-rich) and reducing ($H_2$-rich) primordial Martian atmospheres have been advocated recently (Cannon et al., 2017; Saito and Kuramoto, 2018). Our work offers a path for constraining that atmosphere, recognizing that within the context of terrestrial planetary accretion, primordial atmospheres and hydrospheres are coeval, both deriving from an impact-degassed "steam" atmosphere. The early Martian hydrosphere therefore interacts with – and retains memory of – the primordial atmosphere. Hydrospheric D/H, in particular, yields a proxy for the chemical composition of primordial atmospheres (Pahlevan et al., 2019). This is because: (1) magma-crystal-vapor partitioning predicts that most Martian hydrogen outgasses from the magma ocean and into the primordial atmosphere at the outset of planetary history (Elkins-Tanton, 2008); (2) the greenhouse effect due to a multi-bar $H_2$-rich atmosphere is sufficient to stabilize a water ocean against freezing for the duration of its existence, enabling ocean-atmosphere $H_2O$-$H_2$ isotopic equilibration in early planetary fluid envelopes (Genda and Ikoma, 2008; Pahlevan et al., 2019); (3) deuterium has a strong preference for the water molecule relative to molecular hydrogen, partitioning into primordial hydrospheres at the expense of $H_2$ atmospheres; and (4) the divergent fates of $H_2O$ (retained) and $H_2$ (lost) in primordial planetary atmospheres (see §3.2) due to condensation below the tropopause (the "cold-trap") (Hunten, 1993) implies that the hydrosphere – and its isotopic composition – can be preserved even after the primordial atmosphere is lost via hydrodynamic escape and/or impact erosion. These considerations suggest that the early Martian D/H enrichment yields an oxybarometer reflecting the $H_2/H_2O$ of primordial outgassing and – with an empirically determined $H_2O$ inventory from geomorphology (Di Achille and Hynek, 2010) – provides a barometer for primordial $pH_2$ (>10 bars) (§3.3). In this section, we discuss the removal to space of the primordial atmosphere (§4.1) the subsequent D/H evolution of the Martian hydrosphere (§4.2), and the connection to the redox state of silicate Mars (§4.3).

*4.1. Mechanisms and timescales for primordial atmosphere removal*
Despite the stability of the atmosphere against rapid hydrodynamic expansion powered by the ambient heat of the magma ocean (see supplementary §A1), mechanisms operative in the early Solar System can eliminate the primordial Martian atmosphere on geologic (>Myr) timescales (Lammer et al., 2013).

One such mechanism is extreme ultraviolet (EUV)-powered hydrodynamic escape, a mechanism that relies on the deposition of high-energy photons (e.g., $\lambda$<120 nm) in the upper planetary atmosphere. This mechanism partially ionizes molecules high in the atmosphere and heats the gas, resulting in a lower value of the escape parameter than that prevailing in the troposphere (Equation A1). For a sufficiently intense flux of EUV radiation from the young Sun ($\gtrsim$10-20 × the modern flux), this process induces hydrodynamic outflow from an $H_2O$-$CO_2$-rich (Erkaev et al., 2014; Tian et al., 2009) or $H_2$-CO-$CH_4$-rich (Yoshida and Kuramoto, 2020) Martian atmosphere, at a very approximate rate of ≈1 bar/Myr.

A second mechanism is impact erosion, also called atmospheric cratering, a process in which high-velocity solid-body impactors deposit energy into a planetary atmosphere, accelerating a fraction of the gas onto unbound trajectories (Cameron, 1983). Loss by this mechanism occurs in distinct regimes. For large ("giant") impacts, global shocks propagate through the body of the planet, inducing global ground motions that shock the atmosphere globally, including far from the impact site (Genda and Abe, 2003). The largest impact on Mars of which we have a record is the putative Borealis basin forming impact, which is thought to involve an impactor with a radius at most ≈700-1200 km, i.e., a few percent of a Martian mass (Marinova et al., 2008; Nimmo et al., 2008). Newly developed scaling laws (Schlichting et al., 2015) predict that such an impact removes <3% of the primordial Martian atmosphere. For smaller impacts, the global loss of gas is negligible, and atmospheric escape is limited to regional loss, with the air mass above the tangent plane to the impact serving as an upper limit (Melosh and Vickery, 1989). Although the cumulative magnitude of such impact erosion depends on the number of impactors that are capable of ejecting atmosphere, and hence, on the size distribution of planetesimal populations (Schlichting et al., 2015), the minimum impactor size for ejection depends on the density and scale height of the atmosphere. For this reason, hot and massive $H_2$-rich atmospheres may exhibit a self-stabilizing character with respect to impact erosion, until they are depleted by other mechanisms (e.g., hydrodynamic escape) and made less dense and more susceptible to losses via impact-induced accelerations by smaller (i.e., more numerous) impactors.

Although the relative roles of hydrodynamic escape and impact erosion in removing primordial atmosphere are uncertain, there is likely an interplay between these processes in early Martian history. Moreover, both mechanisms imply atmospheric lifetimes of millions of years, sufficiently long to produce ocean-atmosphere D/H equilibration (Pahlevan et al., 2019), justifying the use of equilibrium isotopic partitioning to make inferences (see §3.3).

*4.2. Late D/H evolution of the Martian hydrosphere*
Because Mars lacks plate tectonics, the initial D/H of outgassed Martian hydrogen is expected to be preserved in magma ocean crystallization products, i.e., the Martian mantle. Estimates of the D/H value of this mantle reservoir cluster around the terrestrial value (Greenwood et al., 2018; Hallis et al., 2012; Peslier et al., 2019; Usui et al., 2012). Most hydrogen loss apparently occurs in the first ≈500 Myr, as shown by the D/H enrichment (≈2-3× SMOW) by the time of ALH84001 carbonate formation at 3.9-4.1 Gya (Borg et al., 1999; Greenwood et al., 2008). Reducing conditions during magma ocean outgassing naturally reproduces this early D/H enrichment because most Martian hydrogen initially appears as $H_2$, which isotopically equilibrates with – and transfers its deuterium to – surface $H_2O$ before escape (Fig. 5). This model leaves unanswered the question of how the hydrosphere and atmosphere isotopically evolved later, from the early (≈2-3×) D/H enrichment to that of the modern atmosphere (≈6×) (Fig. 1). The problem is that the modern hydrogen escape rate is several orders of magnitude too low to isotopically enrich the massive early hydrosphere, which fluvial geomorphology indicates was >500 m GEL (Owen et al., 1988; Yung et al., 1988).

Motivated by the requirement of fractionating an initially massive hydrosphere, the hydrogen escape rate was recently re-evaluated for a suite of plausible ancient Martian atmospheres (Scheller et al., 2021). These authors conclude that the escape rate deriving from the photodissociation of water is insufficient and argue that hydrospheric D/H enrichment is suppressed until irreversible crustal hydration reduces the surface water inventory, so that modest hydrogen escape fluxes then enrich the D/H of a smaller remnant reservoir. Such a model is compatible with – and would post-date – the scenario we have presented. Whereas (Scheller et al., 2021) focus on the "late" D/H evolution (4 Ga-present) and consider $CO_2$-rich atmospheres with $H_2$ constituting a minor species ($xH_2=10^{-3}$), we are interested in "early" D/H evolution (4.5-4 Ga) and consider $H_2$-rich atmospheres ($xH_2=0.9-1$). If the conclusions reached in (Scheller et al., 2021) are correct, and a massive early hydrosphere cannot become deuterium-enriched via hydrogen escape in a $CO_2$-rich atmosphere, then early deuterium enrichment of Mars's hydrosphere relative to the mantle (Fig. 1) may *require* a primordial $H_2$ inventory to accommodate the inferred high early escape flux, a conclusion compatible with the model we describe.

*4.3. Relationship to the redox state of silicate Mars*

Recent estimates of the timescales for magma ocean-atmosphere equilibration suggest that this process is rapid relative to evolutionary timescales (Hamano et al., 2013; Pahlevan et al., 2019). To the extent that magmatic and atmospheric reservoirs are able to equilibrate, the oxygen fugacity characterizing their composition at the interface is expected to be equal. Any difference in the oxidation state of the primordial atmosphere and silicate Mars, therefore, requires explanation. The inferred $fO_2$ of magma ocean-atmosphere equilibration from D/H oxybarometry is $\Delta IW<-1$ (see §3.3). Such a low $fO_2$ is consistent with conditions prevailing during Martian core formation (Brennan et al., 2022) and expected for an atmosphere that equilibrates with a magma ocean hosting convectively suspended metallic droplets. By contrast, the $fO_2$ inferred for magmas deriving from the Martian mantle are substantially higher ($\Delta IW=0$ to $+4$) (Wadhwa, 2008). Moreover, there is uncertainty about the oxidation state of mantle domains, due to the confounding effects of auto-oxidation due decompression, fractional crystallization, and volatile degassing, processes that can alter the redox state of magmas relative to sources (Castle and Herd, 2017; Nakada et al., 2020; Righter et al., 2013). Nevertheless, it appears that some process has oxidized the Martian mantle from an early reduced state in equilibrium with metallic Fe during accretion to a moderately oxidized modern state with measurable amounts of ferric iron.

Two magma ocean processes have been identified that can potentially oxidize silicate Mars and reconcile a reducing primordial atmosphere with a moderately oxidized mantle. First, outgassing of magmatic water (dissolved as $H_2O$ and OH) as gaseous $H_2$ can leave behind an oxidized magma (Sharp et al., 2013). Second, although Mars has internal pressures too low for bridgmanite stability (Stähler et al., 2021), P-T conditions in a magma ocean may nevertheless be sufficiently high to permit magma oxidation due to Fe disproportionation accompanying liquid metal-liquid silicate equilibration ($3Fe^{2+} \rightarrow 2Fe^{3+} + Fe^0$) during core formation (Hirschmann, 2022). For either process to reconcile the inference of a reducing primordial atmosphere ($\Delta IW<-1$ §3.3) with an oxidized modern mantle, it must have been operative after last equilibration of the magma ocean and atmosphere. Evidence for magma ocean-atmosphere disequilibrium has recently emerged from measurements of krypton isotopes (Péron and Mukhopadhyay, 2022). Whether divergent magma ocean-atmosphere redox evolution can quantitatively reconcile a reducing atmosphere with a more oxidized mantle must be subject to future study.

**5. Conclusions**

The hydrogen isotopic composition of the Martian crust-mantle system constrains the evolution of Mars's primordial hydrosphere and atmosphere. Empirical isotopic evidence points to volatile-refractory element fractionation on newly-formed Mars, a process commonly associated with magma ocean outgassing (Marty and Marti, 2002). It is generally thought by modelers that most Martian hydrogen and carbon outgasses primordially (Elkins-Tanton, 2008). Because Mars lacks plate tectonics, the mantle records the D/H of water initially dissolved in the magma ocean. By contrast, ancient alteration minerals in the crust record a ≈2-3× D/H enrichment in the hydrosphere, an early-established global reservoir that reflects preferential escape of protium relative to deuterium from the primordial Martian atmosphere.

The significant conclusions from this study include:
1. Greenhouse warming by $1-10^2$ bars of $H_2$ in the primordial Martian atmosphere in co-existence with water oceans produces global-average surface temperatures of 290-560 K.
2. An efficient condensation cold-trap retains water vapor in the lower atmosphere of $H_2$-rich atmospheres, justifying the divergent fate approximation for H-bearing molecules, which are either retained by condensation ($H_2O$) or transported to the upper atmosphere and lost via escape ($H_2$).
3. The strong preference of deuterium for the water molecule at equilibrium makes hydrospheric D/H a reflection of the outgassed atmospheric $H_2/H_2O$ ratio and an oxybarometer for the last atmospheric equilibration with the magma ocean.
4. The early-formed ≈2-3× Martian crust-mantle D/H offset is produced by primordial outgassing of the atmosphere at reducing conditions ($\Delta IW<-1$), implying that most Martian hydrogen initially appears in reducing form, and yields a barometer for the initial Martian $H_2$ inventory ($pH_2>10$ bars). This greenhouse gas inventory implies a warm ($T_S>400$ K) initial climate for post-magma-ocean Mars.
5. Reducing conditions in the primordial Martian atmosphere inferred via D/H imply that Martian carbon initially appears in reducing form (e.g., CO or $CH_4$). Such an initial condition can be reconciled with the modern $CO_2$ atmosphere of Mars via preferential $H_2O$ retention and $H_2$ escape (e.g., $CO+H_2O \rightarrow CO_2+H_2$). The retention of water vapor via condensation gradually confers on $H_2O$ the role of the oxidant of the Martian surface environment, a role it maintains to the present. An early CO-rich Martian atmosphere is of interest because such an environment is conducive to prebiotic chemistry and has been implicated in the origin of life in the Solar System (Kasting, 2014) and may have left an observable record that may be detectable by the current generation of Mars spacecraft.

**Acknowledgements:** K.P. acknowledges support from a grant from the W.M. Keck Foundation (PI: Peter Buseck) and NASA's Emerging Worlds and Solar System Workings programs (PI: Kaveh Pahlevan). The authors thank Bethany Ehlmann, Edwin Kite, and Robin Wordsworth for comments that helped to improve an early draft of the manuscript.

**Supplementary Information**
**A1. Stability of primordial atmospheres against thermally-driven hydrodynamic expansion**
The stability of a planetary atmosphere against thermally-driven hydrodynamic expansion ("blow-off") is assessed by calculating the escape parameter:

$$\lambda = GM\mu/rk_BT \tag{A1}$$

with G the gravitational constant, M the planetary mass, $\mu$ the mean molecular weight of the atmospheric gas, $k_B$ Boltzmann's constant, T the ambient atmospheric temperature, and r the planetocentric distance. Sufficiently high values of the escape parameter imply hydrodynamic stability against thermal expansion (and very slow atom-by-atom Jeans escape described by the kinetic theory of gases) whereas sufficiently low values imply wholesale hydrodynamic expansion and blow-off of the planetary atmosphere to space, requiring a fluid dynamical description. The critical value of the escape parameter marking hydrodynamic blow-off onset is 2-3 and 2.4-3.6 for monoatomic and diatomic gases, respectively (Volkov et al., 2011).

We seek to determine the conditions under which an impact-degassed $H_2$-CO-rich atmosphere is stable with respect to such thermally-driven hydrodynamic blow-off. For this reason, we evaluate the escape parameter at the planetary surface for several $H_2$-CO gas mixtures as a function of surface temperature. A pure $H_2$ atmosphere ($\mu$=2) near the Martian surface (r=$R_p$) at magma ocean temperatures ($T_s$=1,500-3,000 K) would have an escape parameter ($\lambda$=1-2) below the threshold for stability ($\lambda \approx$2-3) and would be subject to rapid dissipation via thermally-driven hydrodynamic flow. However, even modest additions of CO (e.g., CO:$H_2$ of 1:3-1:2) significantly increase the mean molecular weight of the gas mixture ($\mu$=8.5-10.7), resulting in hydrodynamic *stability* of the atmosphere with respect to thermally-driven expansion ($\lambda$=4.25-10.7). Carbon in the form of $CH_4$ serves a similar role to CO (Saito and Kuramoto, 2018). In this context, it is noteworthy that the abundances of hydrogen and carbon in carbonaceous chondrites – the presumed source of Martian volatiles – are comparable by number (Kuramoto, 1997; Lodders and Fegley, 1997) and the magma solubility of carbon is substantially lower than that of water (Gaillard et al., 2022), promoting the partitioning of carbon into gas and facilitating its role as a stabilizer of the primordial Martian atmosphere. Such hydrodynamic stability with respect to thermally-powered expansion implies that these atmospheres can persist for geologically significant time periods (>Myr).

**A2. Primordial Martian hydrogen inventory**
In addition to the oxygen fugacity of the magma ocean, which determines the $H_2$/$H_2O$ of the outgassed atmosphere, primordial atmosphere models require specification of the total hydrogen inventory. For this purpose, we use two approaches. First, both hydrogen (Hallis et al., 2012; Peslier et al., 2019; Usui et al., 2012) and nitrogen (Marti and Mathew, 2000; Mathew et al., 1998) isotopic abundances inferred for the Martian mantle are broadly chondritic. Indeed, proposed chondritic Martian building blocks include carbonaceous (Lodders and Fegley, 1997) or enstatite (Sanloup et al., 1999) chondrites, which are volatile-rich. Moreover, shock-degassing and impact release of volatiles into the primordial atmosphere is expected during accretion when proto-Mars exceeds one lunar mass (Lange and Ahrens, 1982; Tyburczy et al., 1986). Accordingly, we calculate primordial atmospheric inventories by assuming that volatile abundances of chondritic building blocks are quantitatively released into the Martian atmosphere.

Proposed chondrite mixtures for Mars range from 0.85-0.11-0.04 H-CI-CV (Lodders and Fegley, 1997) to 0.3-0.7 H-EH or 0.55-0.45 H-EH (Sanloup et al., 1999). These chondritic mixtures are consistent with the stable isotopic data (Warren, 2011). The initial inventory of the primordial atmosphere can therefore be constrained by volatile element abundances of chondritic building blocks (Schaefer and Fegley, 2017; Wasson and Kallemeyn, 1988). Restricting the discussion to hydrogen and assuming atmospheric retention during low-velocity collisions characterizing accretion (Melosh and Vickery, 1989), chondritic Martian models are endowed with >100 bars $H_2$ (Table A1). Due to volatile loss from precursor planetesimals (Lichtenberg et al., 2019), partial retention during magma ocean solidification (Elkins-Tanton, 2008), and volatile dissolution into core-bound metals (Hirschmann, 2012), these abundances are considered upper limits to the hydrogen budget of the primordial atmosphere.

**Table A1 – Initial endowments for chondrite mixtures.** Abundances for carbonaceous (Wasson and Kallemeyn, 1988) and ordinary and enstatite (Schaefer and Fegley, 2017) chondrites adopted from literature values. Adopted values in weight percent hydrogen for CV, CI, H, and EH are 0.28, 2.0, 0.046 and 0.13, respectively. Chondritic endowments are calculated as simple mixtures.

| Mixture | H (wt%) | $pH_2$ (bars) | Reference |
|---|---|---|---|
| 0.85-0.11-0.04 H-CV-CI | 0.150 | 247 | (Lodders and Fegley, 1997) |
| 0.3-0.7 H-EH | 0.105 | 173 | (Sanloup et al., 1999) |
| 0.55-0.45 H-EH | 0.084 | 138 | (Sanloup et al., 1999) |

A second approach for constraining Martian initial $H_2$ abundances is empirical. Models predict that most water dissolved in a magma ocean will outgas upon solidification (Elkins-Tanton, 2008). Indeed, recent inferences suggest that shorelines from a putative ocean in Borealis Basin predate Tharsis (Citron et al., 2018), consistent with a mainly *primordial* outgassing scenario producing a large early endowment that gradually depletes over time (Scheller et al., 2021). Hence, scaling the $H_2/H_2O$ values from magma ocean outgassing (Fig. 2) by the size of the early Martian hydrosphere (≥500 m GEL) (Clifford and Parker, 2001) can constrain initial $H_2$ abundances. In this way, initial $H_2$ abundances can be expressed as a function of the oxygen fugacity of primordial outgassing and is equivalent to 8.8 and 28 bars of pure $H_2$ for outgassing at log$fO_2$=IW-1 and IW-2, respectively. To encompass a plausible range of hydrogen inventories using both approaches, we consider initial hydrogen abundances equivalent to 10-100 bars of pure $H_2$. Constraining the initial H inventory is important because it determines the initial temperatures and duration of existence of the $H_2$-based greenhouse (Figure A1).

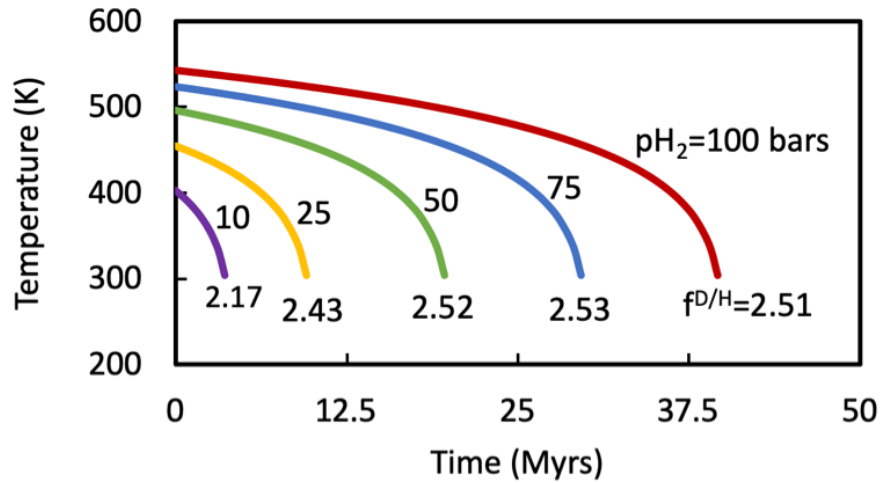

**Fig. A1. Permitted climate scenarios.** The initial climate and the lifetime of the primordial greenhouse depend on the $H_2$ inventory. Extreme ultraviolet (EUV) radiation from the young Sun can power escape on million year timescales. Calculating escape with the energy-limited approximation (Watson et al., 1981) results in a loss rate of ≈2.5 bars/Myrs. In such scenarios, earliest Mars experiences habitable conditions (<100°C) for several million years. The final deuterium-enrichment ($f^{D/H}$) of the hydrosphere (≈2.17-2.51) only weakly depends on the $H_2$ inventory for the full range of endowments here considered ($pH_2$=10-100 bars) and overlaps with the enrichment observed in the Martian isotopic record (≈2-3x, cf. Fig. 1). Such a robust model outcome is evidence for the past presence of an $H_2$ greenhouse on post-magma-ocean Mars.

Although Mars may accrete in the presence of the solar nebula (Dauphas and Pourmand, 2011) and attract nebular hydrogen (Ikoma and Genda, 2006), volatile-rich impactors would produce a high mean molecular weight degassed atmosphere underlying the low molecular weight nebular gas with stable stratification, such that escape of nebular and degassed components would be sequential (Saito and Kuramoto, 2018). Because there is no evidence of a nebular component in the Martian mantle hydrogen reservoir (Usui, 2019), and because we are interested in explaining the D/H *enrichment* due to hydrogen escape, we assume that any nebular hydrogen was lost before exposure of the magma ocean degassed atmosphere to the solar EUV photons and the associated losses. Atmospheres we consider in this work are therefore of purely degassed origin.